\documentclass{article}
\usepackage{spconfa4,amsmath,graphicx}

\usepackage[printonlyused]{acronym}
\usepackage{psfrag}
\usepackage{xcolor}
\usepackage[tight]{subfigure}
\usepackage{pgfplots}
\usetikzlibrary{patterns}

\pgfplotsset{compat=1.3}
\pgfplotsset{
  /pgfplots/xbar legend/.style={
  /pgfplots/legend image code/.code={%
  \draw[##1,/tikz/.cd, bar width=3pt,yshift=-0.1em,bar shift=0pt]
  plot coordinates {(0.8em, 0cm) (0.6em, 1.7*\pgfplotbarwidth)};},
  }
}

\acrodef{ASR}{Automatic Speech Recognition}
\acrodef{DNN}{Deep Neural Network}
\acrodef{DOA}{Direction Of Arrival}
\acrodef{DTFT}{Discrete-Time Fourier Transform}
\acrodef{EARS}{Embodied Audition for RobotS}
\acrodef{FIR}{Finite Impulse Response}
\acrodef{FSU}{Filter-and-Sum Unit}
\acrodef{fwSegSNR}{frequency-weighted segmental Signal-to-Noise Ratio}
\acrodef{GMM}{Gaussian Mixture Model}
\acrodef{HMM}{Hidden Markov Model}
\acrodef{HRTF}{Head-Related Transfer Function}
\acrodef{LS}{Least-Squares}
\acrodef{MSE}{Mean Squared Error}
\acrodef{PFSB}{Polynomial Filter-and-Sum Beamformer}
\acrodef{PLD}{Prototype Look Direction}
\acrodef{PPF}{Polynomial Postfilter}
\acrodef{RIR}{Room Impulse Response}
\acrodef{RLSFI}{Robust Least-Squares Frequency-Invariant}
\acrodef{RLSFIP}{Robust Least-Squares Frequency-Invariant Polynomial}
\acrodef{WNG}{White Noise Gain}
\acrodef{WER}{Word Error Rate}

\newcommand{\bb}[1]{\mathbf{#1}}
\newcommand{\mrm}[1]{\mathrm{#1}}
\DeclareMathOperator*{\argmin}{argmin}

\title{HRTF-based Robust Least-Squares Frequency-Invariant Polynomial Beamforming}
\name{Hendrik Barfuss, Marcel Mueglich, and Walter Kellermann\thanks{The research leading to these results has received funding from the European Union's Seventh Framework Programme (FP7/2007-2013) under grant agreement n$^\mathsf{o}$ 609465.}}

\address{Multimedia Communications and Signal Processing,\\
         Friedrich-Alexander University Erlangen-N\"urnberg\\
         Cauerstr. 7, 91058 Erlangen, Germany \\
{\{hendrik.barfuss,walter.kellermann\}@fau.de, marcel.mueglich@studium.fau.de}}

\begin{document}
\ninept
\maketitle
\begin{abstract} 
In this work, we propose a robust \ac{HRTF}-based polynomial beamformer design which accounts for the influence of a humanoid robot's head on the sound field. In addition, it allows for a flexible steering of our previously proposed robust \ac{HRTF}-based beamformer design.
We evaluate the \ac{HRTF}-based polynomial beamformer design and compare it to the original \ac{HRTF}-based beamformer design by means of signal-independent measures as well as word error rates of an off-the-shelf speech recognition system.
Our results confirm the effectiveness of the polynomial beamformer design, which makes it a promising approach to robust beamforming for robot audition.
\end{abstract}
%

\acresetall

\section{Introduction}
\label{sec:intro}
Spatial filtering techniques are a widely used means to spatially focus on a target source by exploiting spatial information of a wave field which is sampled by several sensors at different positions in space. 

When spatial filtering techniques are applied to a robot audition scenario, i.e., when the microphones are mounted on a humanoid robot's head, the influence of the head on the sound field has to be taken into account by the beamformer design in order to obtain a satisfying spatial filtering performance. To this end, \acp{HRTF}\footnote{In the context of this work, \acp{HRTF} only model the direct propagation path between a source and a microphone mounted on a humanoid robot's head, but no reverberation components.} can be incorporated into the beamformer design as steering vectors, see, e.g., \cite{Pedersen_ICASSP_2004,Maazaoui_EURASIP_2012,Maazaoui_IWSSIP_2012}.
%
In \cite{lnt2009-22}, Mabande et al.~proposed a \ac{RLSFI} beamformer design which allows the user to directly control the tradeoff between the beamformer's spatial selectivity and its robustness. Recently, we extended this design to an \ac{HRTF}-based \ac{RLSFI} beamformer design by following the approach described above \cite{lnt2015-31}. Despite all advantages of the beamformer designs in \cite{lnt2009-22, lnt2015-31}, a clear disadvantage is that whenever the beamformer is steered to another direction, a new optimization problem has to be solved which makes it unattractive for real-time processing. To overcome this limitation, Mabande et al.~proposed a \ac{RLSFIP} beamformer design \cite{lnt2010-46} as extension of \cite{lnt2009-22}, which allows for a flexible steering of the beamformer.

In this work, we extend the \ac{HRTF}-based \ac{RLSFI} beamformer design \cite{lnt2015-31} to the concept of polynomial beamforming in order to allow for a flexible steering of the \ac{HRTF}-based beamformer in a robot audition scenario.

The remainder of this article is structured as follows: In Section~\ref{sec:HRTFbasedRobustPolynomialbeamforming}, the \ac{HRTF}-based \ac{RLSFIP} beamformer design is introduced. Then, an evaluation of the new \ac{HRTF}-based polynomial beamformer design is presented in Section~\ref{sec:Evaluation}. Finally, conclusions and an outlook to future work are given in Section~\ref{sec:Conclusion}.

\section{HRTF-based robust polynomial beamforming}
\label{sec:HRTFbasedRobustPolynomialbeamforming}

\subsection{Concept of polynomial beamforming}
\label{subsec:conceptPolynomialBeamforming}
\begin{figure}[t]
  \centering
  \scriptsize
  \psfrag{x0}[cl][cl]{$x_{0}[k]$}
  \psfrag{x1}[cl][cl]{$x_{1}[k]$}
  \psfrag{xN}[cl][cl]{$x_{N-1}[k]$}
  \psfrag{y0}[cl][cl]{$y_{0}[k]$}   
  \psfrag{y1}[cl][cl]{$y_{1}[k]$}   
  \psfrag{yP}[cl][cl]{$y_{P}[k]$}   
  \psfrag{yD}[cl][cl]{$y_{D}[k]$}
  \psfrag{D}[cl][cl]{$D$}
  \psfrag{w00}[c][c]{$\bb{w}_{0,0}$}
  \psfrag{w10}[c][c]{$\bb{w}_{1,0}$}
  \psfrag{wN0}[c][c]{$\bb{w}_{N-1,0}$} 
  \psfrag{w01}[c][c]{$\bb{w}_{0,1}$}
  \psfrag{w11}[c][c]{$\bb{w}_{1,1}$}
  \psfrag{wN1}[c][c]{$\bb{w}_{N-1,1}$} 
  \psfrag{w0P}[c][c]{$\bb{w}_{0,P}$}
  \psfrag{w1P}[c][c]{$\bb{w}_{1,P}$}
  \psfrag{wNP}[c][c]{$\bb{w}_{N-1,P}$}   
  \psfrag{p1}[cl][cl]{}
  \psfrag{p2}[cl][cl]{}
  \psfrag{pP}[cl][cl]{}
  \psfrag{+}[c][c]{$+$}    
  \psfrag{F}[c][c]{FSUs}
  \psfrag{PPF}[c][c]{PPF}
  \includegraphics[scale = .95]{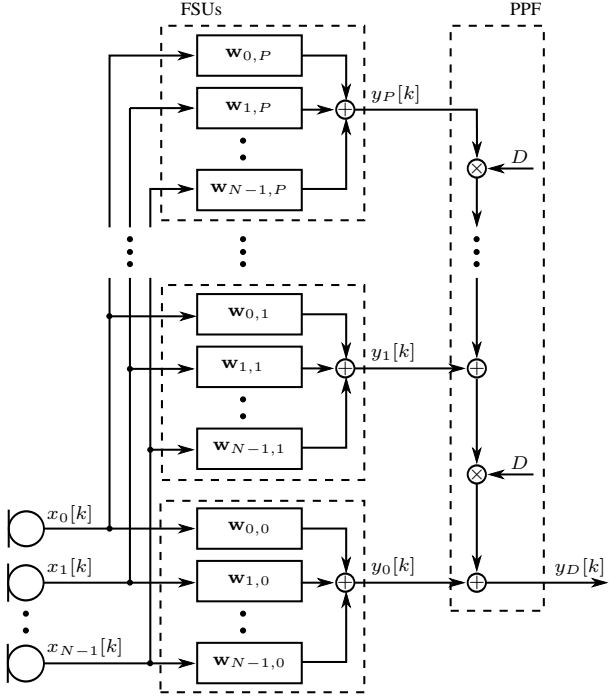}
  \caption{Illustration of a polynomial filter-and-sum beamformer after \cite{lnt2010-46}.}
  \label{fig:polynomialFSB}  
  \vspace{-6mm}
\end{figure}

In Fig.~\ref{fig:polynomialFSB}, the block diagram of a \ac{PFSB}, as presented in \cite{lnt2010-46, Kajala_ICASSP_2001, lnt2014-74}, is illustrated. It consists of a beamforming stage containing $P+1$ \acp{FSU}, followed by a \ac{PPF}. The output signal $y_{p}[k]$ of the $p$-th \ac{FSU} at time instant $k$ is obtained by convolving the microphone signals $x_n[k], \, n \in \{0, \ldots, N-1\}$ with the \ac{FSU}'s \ac{FIR} filters $\bb{w}_{n,p} = [w_{np,0}, \ldots, w_{np,L-1}]^\text{T}$ of length $L$, followed by a summation over all $N$ channels. Operator $(\cdot)^\text{T}$ represents the transpose of vectors or matrices, respectively.
In the \ac{PPF}, the output $y_{D}[k]$ of the \ac{PFSB} is obtained by weighting the output of each \ac{FSU} by a factor $D^{p}$ and summing them up:
\begin{equation}
 y_{D}[k] = y_{0}[k] + Dy_{1}[k] + D^{2}y_{2}[k] + \ldots + D^{P}y_{P}[k].
 \label{eq:y_PFSB}
\end{equation}
Hence, the output signal of each \ac{FSU} can be seen as one coefficient of a polynomial of order $P$ with variable $D$.
The advantage of a \ac{PFSB} is that the steering of the main beam is accomplished by simply changing the scalar value $D$, whereas the \ac{FIR} filters of the \acp{FSU} can be designed beforehand and remain fixed during runtime. A more detailed explanation of how the steering is controlled by $D$ is given in Section~\ref{subsec:HRTFbasedRLSFIPbeamforming}.

The beamformer response of the \ac{PFSB} is given as \cite{lnt2010-46}:
\begin{equation}
  B_{D}(\omega, \phi, \theta) = \sum\limits_{p=0}^{P} D^{p} \sum\limits_{n=0}^{N-1} W_{n,p}(\omega) g_{n}(\omega, \phi, \theta),
  \label{eq:BFResponse_PBF}
\end{equation}
where $W_{n,p}(\omega) = \sum_{l=0}^{L-1} w_{np,l} e^{-j \omega l}$ is the \ac{DTFT} representation of $\mathbf{w}_{n,p}$, and $g_{n}(\omega, \phi, \theta)$ is the sensor response of the $n$-th microphone to a plane wave with frequency $\omega$ traveling in the direction $(\phi, \theta)$. Variables $\phi$ and $\theta$ denote azimuth and elevation angle, and are measured with respect to the positive x-axis and the positive z-axis, respectively, as in \cite{VanTrees:2004}.

\subsection{HRTF-based robust least-squares frequency-invariant polynomial beamforming}
\label{subsec:HRTFbasedRLSFIPbeamforming}
The main goal of the proposed \ac{HRTF}-based \ac{RLSFIP} beamformer design is to jointly approximate $I$ desired beamformer responses $\hat{B}_{D_{i}}(\omega, \phi, \theta)$, each with a different \ac{PLD} $(\phi_{i},\theta_{i}), \, i = 0, \ldots, I-1$, by the actual beamformer response $B_{D_{i}}(\omega, \phi, \theta)$, where $D_{i} = (\phi_{i}-90)/90$, in the \ac{LS} sense. Hence, $D_{i}$ lies in the interval $-1 \leq D_{i} \leq 1$, where, for example, $D=0$ and $D=-1$ steer the main beam towards $\phi=90^\circ$ and $\phi=0^\circ$, respectively. For values of $D$ which do not correspond to one of the \acp{PLD}, the \ac{PPF} will interpolate between them, as expressed in (\ref{eq:y_PFSB}). In this work, we apply polynomial beamforming only in the horizontal dimension. Thus, $D_{i}$ only depends on the azimuth angle, whereas $\theta_{i}$ is constant for all \acp{PLD}. The extension to two-dimensional beam steering is an aspect of future work.
In addition to the \ac{LS} approximation, a distortionless response constraint and a constraint on the \ac{WNG} is imposed on each of the $I$ \acp{PLD}. The approximation is carried out for a discrete set of $Q$ frequencies $\omega_{q}, \, q \in \{0, \ldots, Q-1\}$ and $M$ look directions $(\phi_{m},\theta_{m}), \, m \in \{0, \ldots, M-1\}$ (where, in this work, $\theta_{m}$ remains fixed) in order to obtain a numerical solution. Hence, the optimization problem of the \ac{HRTF}-based \ac{RLSFIP} beamformer design can be expressed as:
\begin{equation}
  \argmin\limits_{\bb{w}_\mrm{f}(\omega_{q})} \sum\limits_{i=0}^{I-1} \Vert \bb{G}(\omega_{q}) \bb{D}_{i} \bb{w}_\mrm{f}(\omega_{q}) - \bb{\hat{b}}_{i} \Vert_{2}^{2},
  \label{eq:OP_LSApproximation}
\end{equation}
subject to $I$ constraints on the corresponding WNG and response in the desired look direction, respectively:
\begin{align}
    \frac{ |\bb{a}^\text{T}_{i}(\omega_{q}) \bb{D}_{i} \bb{w}_\mrm{f}(\omega_{q})|^{2}}{\Vert \bb{D}_{i}\bb{w}_\mrm{f}(\omega_{q}) \Vert_{2}^{2}} & \ge \gamma > 0, \quad  \bb{a}^\text{T}_{i}(\omega_{q}) \bb{D}_{i} \bb{w}_\mrm{f}(\omega_{q}) = 1,\nonumber \\
    & \forall i = 0, \ldots, I-1.  
    \label{eq:OP_Constraints}
\end{align}
where $\displaystyle \bb{\hat{b}}_{i} = [\hat{B}_{D_i}(\phi_{0},\theta_{0}), \ldots, \hat{B}_{D_i}(\phi_{M-1},\theta_{M-1})]^\text{T}$ is a vector of dimension $M \times 1$ containing the $i$-th desired response for all $M$ angles, matrix $\displaystyle [\bb{G}(\omega_{q})]_{mn} = g_{n}(\omega_{q}, \phi_{m}, \theta_{m})$, vector $\bb{a}_{i}(\omega_{q})=[g_{0}(\omega_{q}, \phi_{i}, \theta_{i}), \ldots, g_{N-1}(\omega_{q}, \phi_{i}, \theta_{i})]^\text{T}$ is the steering vector which contains the sensor responses for the $i$-th \ac{PLD} $(\phi_{i},\theta_{i})$, and vector $\displaystyle \bb{w}_\mathrm{f}(\omega_{q}) = [W_{0,0}(\omega_{q}), \ldots,$ $W_{N-1,P}(\omega_{q})]^\text{T}$ of dimension $N(P+1) \times 1$ contains all filter coefficients. Furthermore, $\bb{D}_{i}=\bb{I}_{N} \otimes [D^{0}_{i}, \ldots, D^{P}_{i}]$ is an $N \times N(P+1)$ matrix, where $\bb{I}_{N}$ is an $N \times N$ identity matrix and $\otimes$ denotes the Kronecker product. Operator $\Vert \cdot \Vert_{2}$ denotes the Euclidean norm of a vector. 
%
The optimization problem in (\ref{eq:OP_LSApproximation}), (\ref{eq:OP_Constraints}) can be interpreted as follows: Equation (\ref{eq:OP_LSApproximation}) describes the \ac{LS} approximation of the $I$ desired responses $\hat{B}_{D_{i}}(\omega_{q}, \phi_{m}, \theta_{m})$ by the actual beamformer response. 
The first part of (\ref{eq:OP_Constraints}) represents the \ac{WNG} constraint which is imposed on each of the $I$ \acp{PLD}. $\gamma$ is the lower bound on the \ac{WNG} and has to be defined by the user. Hence, the user has the possibility to directly control the beamformer's robustness against small random errors like sensor mismatch or position errors of the microphones. The second part of (\ref{eq:OP_Constraints}) ensures a distortionless beamformer response for each of the $I$ \acp{PLD}.

As in \cite{lnt2015-31}, we include measured \acp{HRTF} in (\ref{eq:OP_LSApproximation}) and (\ref{eq:OP_Constraints}) instead of the free-field-based steering vectors (which are only based on the microphone positions and the look directions). By doing this, the beamformer design can account for the influence of the humanoid robot's head on the sound field which would not be the case if we used free-field-based steering vectors as in \cite{lnt2010-46}. The sensor responses are given as $g_{n}(\omega_{q}, \phi_{m}, \theta_{m}) = h_{mn}(\omega_{q})$,
where $h_{mn}(\omega_{q})$ is the \ac{HRTF} modeling the propagation between the $m$-th source position and the $n$-th microphone, mounted at the humanoid robot's head, at frequency $\omega_{q}$. Matrix $\bb{G}(\omega_{q})$ consists of all \acp{HRTF} between the $M$ look directions and the $N$ microphones, and $\bb{a}_{i}(\omega_{q})$ contains the \acp{HRTF} corresponding to the $i$-th \ac{PLD}.

The optimization problem has to be solved for each frequency $\omega_{q}$ separately. We use the same desired response for all frequencies for the design of the polynomial beamformer, which is indicated by the frequency-independent entries of $\hat{\bb{b}}_{i}$ \cite{lnt2009-22, lnt2015-31,lnt2010-46}. The optimization problem in (\ref{eq:OP_LSApproximation}), (\ref{eq:OP_Constraints}) is formulated as a convex optimization problem \cite{lnt2010-46} and we use \textrm{CVX}, a package for specifying and solving convex programs in \textrm{Matlab} \cite{cvx}, to solve it. After the optimum filter weights at each frequency $\omega_{q}$ have been found, \ac{FIR} filters of length $L$ are obtained by \ac{FIR} approximation, see, e.g., \cite{Oppenheim:1999:DSP}, of the optimum filter weights using the \textrm{fir2} method provided by \textrm{Matlab} \cite{fir2}.

\section{Evaluation}
\label{sec:Evaluation}
In the following, we evaluate the proposed \ac{HRTF}-based \ac{RLSFIP} beamformer design and compare it to the \ac{HRTF}-based \ac{RLSFI} beamformer design proposed in \cite{lnt2015-31}. At first, the experimental setup is introduced. Then, the two beamformer designs are compared with respect to their approximation errors of the desired beamformer response. Eventually, the signal enhancement performance is evaluated in terms of \acp{WER} of an \ac{ASR} system.

\subsection{Setup and parameters}
\label{subsec:SetupAndParameters}
\begin{figure}
  \subfigure[Microphone positions.]{    
    \includegraphics[width = 4cm]{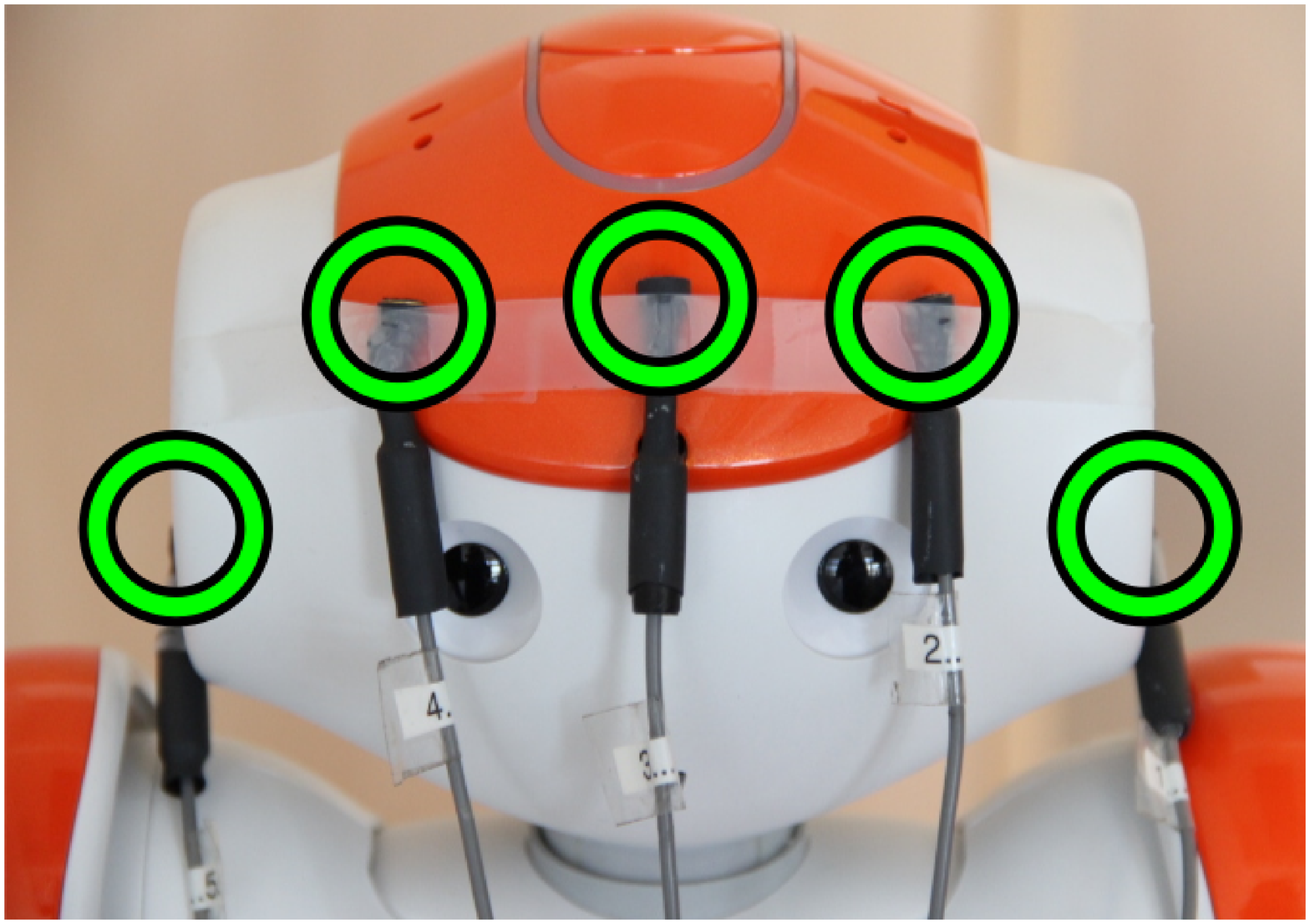}
    \label{fig:setup_headArray}
  }
  \hfill
  \subfigure[Source positions.]{
    \psfrag{inta}[cl][cl]{\parbox[t]{2cm}{\scriptsize \color{red} 1) $\phi_\mrm{int}=70^\circ$}}
    \psfrag{intb}[cl][cl]{\parbox[t]{2cm}{\scriptsize \color{red} 2) $\phi_\mrm{int}=170^\circ$}}
    \psfrag{t}[c][c]{\scriptsize \color{green} target}
    \psfrag{30}[c][c]{\scriptsize $30^\circ$}
    \psfrag{d}[cr][cr]{\scriptsize $1.1\, \text{m}$}
    \psfrag{x1}[cl][cl]{\scriptsize $x_{N-1}$}
    \psfrag{x0}[cl][cl]{\scriptsize $x_0$}
    \includegraphics[width = 3.5cm]{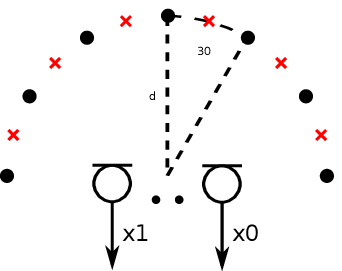}
    \label{fig:evaluation_scenarios}
  } 
  \vspace{-2mm}
  \caption{Illustration of the employed microphone positions (green circles) at the humanoid robot's head and the source positions of the two-speaker scenario.}
  \vspace{-4mm}
\end{figure}

%
The evaluated beamformers were designed for the five-microphone robot head array in Fig.~\ref{fig:setup_headArray}, using a \ac{WNG} constraint of $\gamma_\text{dB}=-20\text{dB}$ and a filter length of $L=1024$.
For the design of the polynomial beamformer, we used $I=5$ \acp{PLD} $\phi_{i} \in \{0^\circ, 45^\circ, 90^\circ, 135^\circ,$ $180^\circ\}$ and a \ac{PPF} of order $P=4$.
The set of \acp{HRTF} which is required for the \ac{HRTF}-based beamformer design was measured in a low-reverberation chamber ($T_{60} \approx 50$ms) using maximum-length sequences, see, e.g., \cite{Schroeder_JASA_1979,Holters_DAF_2009}.
The \acp{HRTF} were measured for the same five-microphone array shown in Fig.~\ref{fig:setup_headArray} for a robot-loudspeaker distance of $1.1$m. The loudspeaker was at an elevation angle of $\theta = 56.4^\circ$ with respect to the robot. We chose this setup to simulate a taller human interacting with the NAO robot which is of height $0.57 \, \text{m}$. The measurements were carried out for the robot looking towards broadside $(\phi,\theta)=(90^\circ,90^\circ)$.


\subsection{Evaluation of HRTF-based polynomial beamformer design}
\label{subsec:EsperimentalResults}
In this section, we investigate how well the desired beamformer response $\hat{B}_{D_{i}}(\phi, \theta)$ is approximated by the beamformer response of either the \ac{HRTF}-based \ac{RLSFI} or the \ac{HRTF}-based \ac{RLSFIP} beamformer.
Ideally, the polynomial beamformer should be as good as the \ac{RLSFI} beamformer in the best case, because it approximates the latter, i.e., the performance of both beamformers should be similar when steered towards one of the $I$ \acp{PLD}.

\begin{figure}[t]
\subfigure{
  \hspace{8mm}
  \begin{tikzpicture}[scale=1,trim axis left]
  \node at (-0.975,1.575) {\scriptsize (a)};
  \node at (6.725,1.875) {\scriptsize $\text{BP}_\text{dB}/\text{dB}$};    
    \begin{axis}[
      label style = {font=\scriptsize},
      tick label style = {font=\tiny},   
      ylabel style={yshift=-1mm}, 
      width=8.91cm,height=3.25cm,grid=major,grid style = {dotted,black},  		
      axis on top, 	
      enlargelimits=false,
      xmin=300, xmax=5000, ymin=0, ymax=180,
      xtick={300,1000,2000,3000,4000,5000},
      xticklabels={\empty}, 
      ytick={0,45,90,135,180},
      ylabel={$\phi/\circ\,\rightarrow$},
      colorbar horizontal, colormap/jet, 
      colorbar style={
	at={(0,1.15)}, anchor=north west, font=\tiny, width=6cm, height=0.15cm, xticklabel pos=upper
      },
      point meta min=-40, point meta max=0]
      \addplot graphics [xmin=270, xmax=5035, ymin=0, ymax=185] {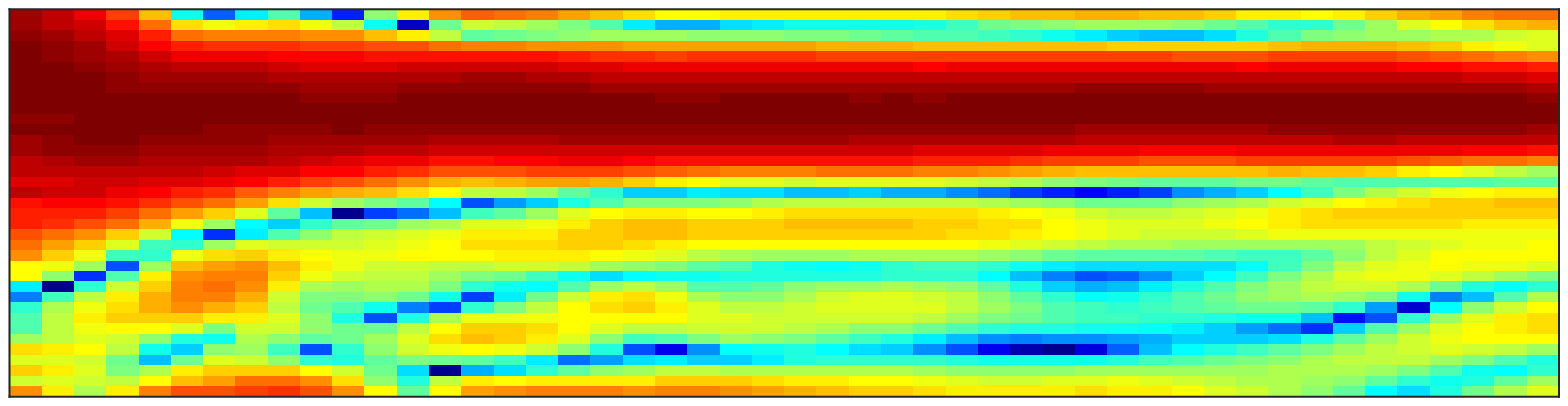};
    \end{axis}
  \end{tikzpicture}
  }
  \\[-4mm]       
  \subfigure{	
  \hspace{8mm}
  \begin{tikzpicture}[scale=1,trim axis left]
  \node at (-0.975,1.575) {\scriptsize (b)};
    \begin{axis}[
      label style = {font=\scriptsize},
      tick label style = {font=\tiny},   
      ylabel style={yshift=-1mm},  	 
      width=8.91cm,height=3.25cm,grid=major,grid style = {dotted,black},  		
      axis on top, 	
      enlargelimits=false,
      xmin=300, xmax=5000, ymin=0, ymax=180,
      xtick={300,1000,2000,3000,4000,5000},
      xticklabels={\empty},    
      ytick={0,45,90,135,180},
      ylabel={$\phi/\circ\,\rightarrow$}]
      \addplot graphics [xmin=270, xmax=5035, ymin=0, ymax=185] {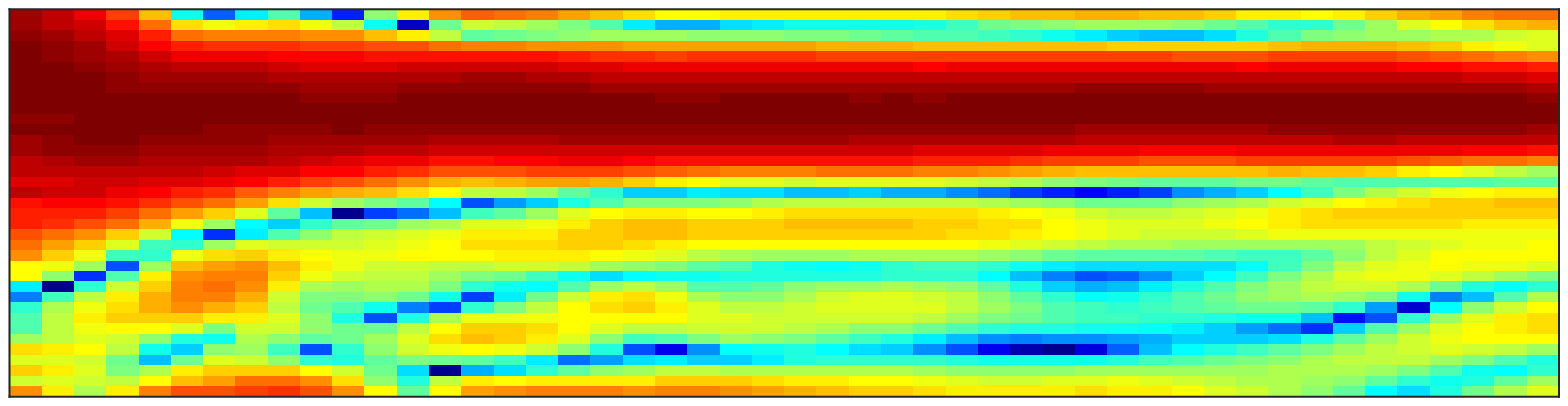};
    \end{axis}
  \end{tikzpicture} 
  }\\[-3.5mm]
  \subfigure{  
  \hspace{8mm}
    \begin{tikzpicture}[scale=1,trim axis left]
    \node at (-0.975,1.575) {\scriptsize (c)};
    \begin{axis}[
      label style = {font=\scriptsize},
      tick label style = {font=\tiny},
      ylabel style={yshift=-2mm},  	
      legend style={font=\scriptsize, yshift=0.25mm, at={(.515,0.97)}},
      legend columns = -1,    
      width=8.91cm,height=3.25cm,grid=major,grid style = {dotted,black},
      xtick={300,1000,2000,3000,4000,5000},
      xticklabels={$300$,$1000$,$2000$,$3000$,$4000$,$5000$},
      xlabel={$f/\text{Hz} \, \rightarrow$},
      ytick={-20, -15, -10, -5, 0},  
      ylabel={$\text{WNG}/\text{dB}\,\rightarrow$},
      ymin=-22.5, ymax=2.5, xmin=300,xmax=5000]   
      \addplot[thick,blue,solid] table [x index=0, y index=1]{WNG_RLSFI_hrtf_nmic_5_az_135_WNGlim_-20.dat}; \addlegendentry{RLSFI$\,\,\,$};
      \addplot[thick,red,dashed] table [x index=0, y index=1]{WNG_RLSFIP_hrtf_nmic_5_az_135_WNGlim_-20.dat}; \addlegendentry{RLSFIP};	
    \end{axis}       
  \end{tikzpicture}    
  }
  \vspace{-8mm}
  \caption{Illustration of beampatterns of (a) the \ac{HRTF}-based \ac{RLSFI} beamformer and (b) the \ac{HRTF}-based \ac{RLSFIP} beamformer when the polynomial beamformer's look direction coincides with a \ac{PLD}. The beamformers were designed for the five-microphone robot head array in Fig.~\ref{fig:setup_headArray} with look direction $(\phi_\text{ld}, \theta_\text{ld})=(135^\circ, 56.4^\circ)$ and \ac{WNG} constraint $\gamma_\text{dB}=-20\, \text{dB}$. The resulting \ac{WNG} is illustrated in Subfigure (c).}
  \label{fig:designexample_prototpyelookDirection}
  \vspace{-6mm}
\end{figure}

Fig.~\ref{fig:designexample_prototpyelookDirection} shows the beampatterns of the \ac{HRTF}-based \ac{RLSFI} beamformer and of the \ac{HRTF}-based \ac{RLSFIP} beamformer in Figs~\ref{fig:designexample_prototpyelookDirection}(a) and \ref{fig:designexample_prototpyelookDirection}(b), respectively, steered towards $(\phi_\text{ld}, \theta_\text{ld})=(135^\circ, 56.4^\circ)$. The resulting \ac{WNG} of both beamformer designs is shown in Fig.~\ref{fig:designexample_prototpyelookDirection}(c). Please note that the beampatterns were computed with \acp{HRTF} modeling the acoustic system. Thus, they effectively show the transfer function between source position and beamformer output. A comparison of the beampatterns of the \ac{HRTF}- and free-field-based \ac{RLSFI} beamformer can be found in \cite{lnt2015-31}, illustrating the effect of the humanoid robot's head as scatterer on the sound field.
%
From Fig.~\ref{fig:designexample_prototpyelookDirection} it can be seen that the beampatterns of both beamformers look almost identical. This is because the actual look direction of the beamformers is equal to one of the five \acp{PLD} of the polynomial beamformer design. 
One can also see that the \ac{WNG} is successfully constrained for both beamformer designs across the entire frequency range (with some slight deviations due to the \ac{FIR} approximation with finite filter length). Comparison of Figs~\ref{fig:designexample_prototpyelookDirection}(a) and \ref{fig:designexample_prototpyelookDirection}(b) with Fig~\ref{fig:designexample_prototpyelookDirection}(c) reveals that the beamformer's main beam broadens when the \ac{WNG} reaches its lower bound. This points to the tradeoff between robustness and spatial selectivity which the user can control via $\gamma$ in (\ref{eq:OP_Constraints}).

\begin{figure}[t]
\subfigure{
  \hspace{8mm}
  \begin{tikzpicture}[scale=1,trim axis left]
  \node at (-0.975,1.575) {\scriptsize (a)};
  \node at (6.725,1.875) {\scriptsize $\text{BP}_\text{dB}/\text{dB}$};    
    \begin{axis}[
      label style = {font=\scriptsize},
      tick label style = {font=\tiny},
      ylabel style={yshift=-1mm}, 
      width=8.91cm,height=3.25cm,grid=major,grid style = {dotted,black},  		
      axis on top, 	
      enlargelimits=false,
      xmin=300, xmax=5000, ymin=0, ymax=180,
      xtick={300,1000,2000,3000,4000,5000},
      xticklabels={\empty}, 
      ytick={0,45,90,135,180},
      ylabel={$\phi/\circ\,\rightarrow$},
      colorbar horizontal, colormap/jet, 
      colorbar style={
	at={(0,1.15)}, anchor=north west, font=\tiny, width=6cm, height=0.15cm, xticklabel pos=upper
      },
      point meta min=-40, point meta max=0]
      \addplot graphics [xmin=270, xmax=5035, ymin=0, ymax=185] {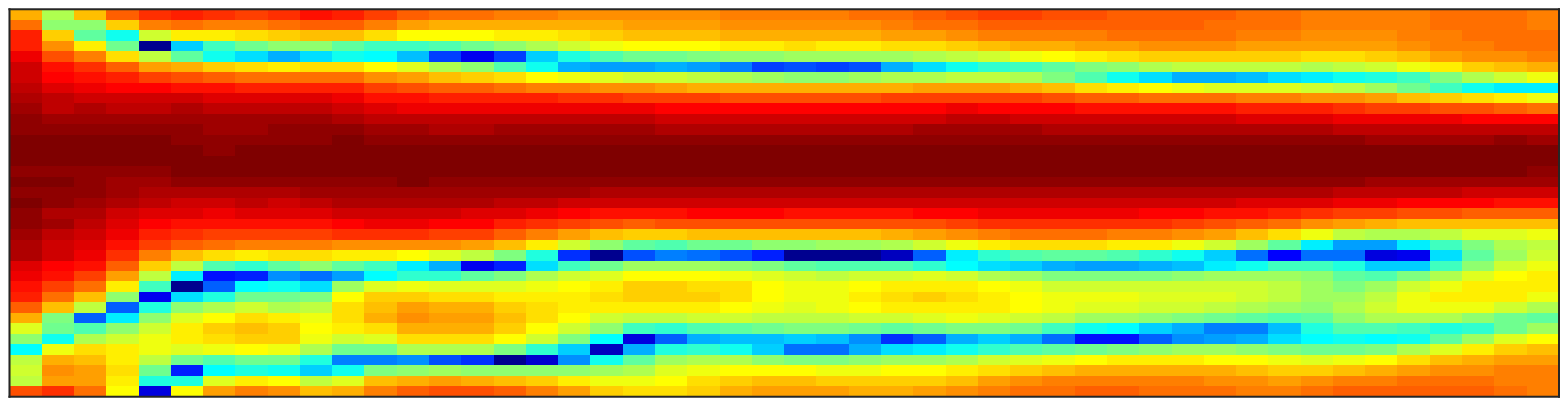};
    \end{axis}
  \end{tikzpicture}
  }
  \\[-4mm]       
  \subfigure{	
  \hspace{8mm}
  \begin{tikzpicture}[scale=1,trim axis left]
  \node at (-0.975,1.575) {\scriptsize (b)};
    \begin{axis}[
      label style = {font=\scriptsize},
      tick label style = {font=\tiny},   
      ylabel style={yshift=-1mm},  	 
      width=8.91cm,height=3.25cm,grid=major,grid style = {dotted,black},  		
      axis on top, 	
      enlargelimits=false,
      xmin=300, xmax=5000, ymin=0, ymax=180,
      xtick={300,1000,2000,3000,4000,5000},
      xticklabels={\empty},    
      ytick={0,45,90,135,180},
      ylabel={$\phi/\circ\,\rightarrow$}]
      \addplot graphics [xmin=270, xmax=5035, ymin=0, ymax=185] {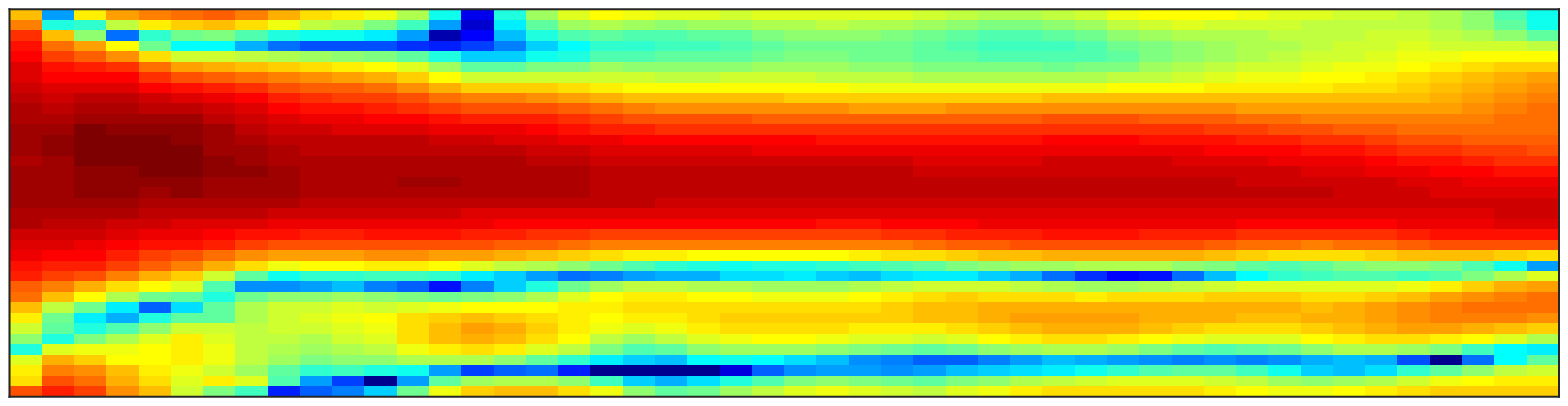};
    \end{axis}
  \end{tikzpicture} 
  }\\[-3.5mm]
  \subfigure{  
  \hspace{8mm}
    \begin{tikzpicture}[scale=1,trim axis left]
    \node at (-0.975,1.575) {\scriptsize (c)};
    \begin{axis}[
      label style = {font=\scriptsize},
      tick label style = {font=\tiny},
      ylabel style={yshift=-2mm},  	
      legend style={font=\scriptsize, yshift=0.25mm, at={(.515,0.97)}},
      legend columns = -1,    
      width=8.91cm,height=3.25cm,grid=major,grid style = {dotted,black},
      xtick={300,1000,2000,3000,4000,5000},
      xticklabels={$300$,$1000$,$2000$,$3000$,$4000$,$5000$},
      xlabel={$f/\text{Hz} \, \rightarrow$},	
      ytick={-20, -15, -10, -5, 0},  
      ylabel={$\text{WNG}/\text{dB} \,\rightarrow$},
      ymin=-22.5, ymax=2.5, xmin=300,xmax=5000]   
      \addplot[thick,blue,solid] table [x index=0, y index=1]{WNG_RLSFI_hrtf_nmic_5_az_110_WNGlim_-20.dat}; \addlegendentry{RLSFI$\,\,\,\,$};
      \addplot[thick,red,dashed] table [x index=0, y index=1]{WNG_RLSFIP_hrtf_nmic_5_az_110_WNGlim_-20.dat}; \addlegendentry{RLSFIP};	
    \end{axis}
  \end{tikzpicture}    
  }
  \vspace{-8.5mm}
  \caption{Illustration of beampatterns of (a) the \ac{HRTF}-based \ac{RLSFI} beamformer and (b) the \ac{HRTF}-based \ac{RLSFIP} beamformer when the polynomial beamformer's look direction does not coincide with one of the \acp{PLD}. The beamformers were designed for the five-microphone robot head array in Fig.~\ref{fig:setup_headArray} with look direction $(\phi_\text{ld}, \theta_\text{ld})=(110^\circ, 56.4^\circ)$ and \ac{WNG} constraint $\gamma_\text{dB}=-20\, \text{dB}$. Subfigure (c) shows the resulting \ac{WNG}.}
  \label{fig:designexample_offPrototpyelookDirection}
\end{figure}

In Fig.~\ref{fig:designexample_offPrototpyelookDirection} the beampatterns of the \ac{HRTF}-based \ac{RLSFI} and \ac{RLSFIP} beamformers are shown for the look direction $(\phi_\text{ld}, \theta_\text{ld})=(110^\circ, 56.4^\circ)$, which lies roughly half-way between two \acp{PLD} and can be expected to exhibit a large deviation from the desired response. The two beampatterns now look different, which is due to the interpolation between the \acp{PLD} by the polynomial beamformer. While for the lower frequencies the two main beams still look similar, the main beam of the polynomial beamformer is degraded for higher frequencies.
Moreover, it can be observed that the polynomial beamformer cannot maintain a distortionless response in the desired look direction across the entire frequency range.
The mismatch between \ac{RLSFI} and \ac{RLSFIP} beamformer also becomes obvious when looking at the \ac{WNG} in Fig.~\ref{fig:designexample_offPrototpyelookDirection}(c). The \ac{WNG} of the \ac{RLSFIP} beamformer is generally lower than that of the \ac{RLSFI} beamformer. In addition, the polynomial beamformer also exhibits a stronger violation of the \ac{WNG} constraint than the \ac{RLSFI} beamformer for $f < 500$Hz.

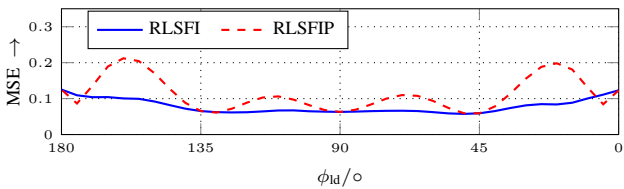
\begin{figure}
  \hspace{8mm}  
    \begin{tikzpicture}[scale=1,trim axis left]
    \begin{axis}[
      label style = {font=\scriptsize},
      tick label style = {font=\tiny},
      ylabel style={yshift=-1mm},  	
      legend style={font=\scriptsize, yshift=0.25mm, at={(.515,0.97)}},
      legend columns = -1,    
      width=8.91cm,height=3.25cm,grid=major,grid style = {dotted,black},
      x dir=reverse,   
      xtick={0, 45, 90, 135, 180},
      xlabel={$\phi_\text{ld}/\circ$},	
      ytick={0, 0.1, 0.2, 0.3},  
      ylabel={$\text{MSE} \,\rightarrow$},
      ymin=0, ymax=0.35, xmin=0,xmax=180]   
      \addplot[thick,blue,solid] table [x index=0, y index=1]{MSE_RLSFI_hrtf_nmic_5_WNGlim_-20.dat}; \addlegendentry{RLSFI$\,\,\,\,$};
      \addplot[thick,red,dashed] table [x index=0, y index=1]{MSE_RLSFIP_hrtf_nmic_5_WNGlim_-20.dat}; \addlegendentry{RLSFIP};	
    \end{axis}       
  \end{tikzpicture}    
  \vspace{-4.5mm}
  \caption{Illustration of the \ac{MSE} (\ref{eq:MSE_BFResponse}) of the \ac{HRTF}-based \ac{RLSFI} (blue curve) and \ac{HRTF}-based \ac{RLSFIP} (red curve) beamformer designs, calculated in steps of five degrees over the entire steering range.}
  \label{fig:MSE_I_5_P_4}
  \vspace{-3mm}
\end{figure}

In the following, we measure the approximation error of the desired response $\hat{B}_{D_\text{ld}}(\phi, \theta)$ for a certain look direction $\phi_\text{ld}$ by the actual beamformer response $B_{D_\text{ld}}(\omega, \phi, \theta)$, where $D_\text{ld} = (\phi_\text{ld}-90)/90$, of either the \ac{RLSFI} or \ac{RLSFIP} beamformer by calculating the \ac{MSE} which is defined as \cite{lnt2014-74}:
\begin{equation}
  \text{MSE}(\phi_\text{ld}) = \sum\limits_{q=0}^{Q-1}\sum\limits_{m=0}^{M-1} \frac{\left( |B_{D_\text{ld}}(\omega_{q}, \phi_{m}, \theta_{m})|-|\hat{B}_{D_\text{ld}}(\phi_{m}, \theta_{m})| \right)^{2}}{Q \cdot M}.
  \label{eq:MSE_BFResponse}
\end{equation}
Fig.~\ref{fig:MSE_I_5_P_4} depicts the \ac{MSE} of the \ac{HRTF}-based \ac{RLSFI} and \ac{RLSFIP} beamformer designs, calculated in steps of five degrees over the entire steering range $0^\circ \leq \phi_\text{ld} \leq 180^\circ$.
When steered to one of the five \acp{PLD}, i.e., when $\phi_\text{ld}=\phi_{i}$, the \ac{RLSFIP} beamformer design yields a similar \ac{MSE} as the \ac{RLSFI} beamformer design. In between those \acp{PLD}, the \ac{MSE} of the polynomial beamformer design is usually larger than that of the \ac{RLSFI} beamformer design. The lower \ac{MSE} of the polynomial beamformer for $\phi_\text{ld} \in \{5^\circ, 175^\circ\}$ may be explained by side lobes of the polynomial beamformer which are less pronounced at higher frequencies than those of the \ac{RLSFI} beamformer for these two particular look directions.

\subsection{Evaluation of signal enhancement performance}
\label{subsec:EsperimentalResults}
In this section, we evaluate the overall quality of the enhanced signals at the outputs of the \ac{HRTF}-based \ac{RLSFI} and \ac{RLSFIP} beamformers. In addition, we also evaluate the original free-field-based \ac{RLSFIP} beamformer \cite{lnt2010-46} which assumes free-field propagation of sound waves and, therefore, cannot account for the influence of robot's head on the sound field. To this end, we use \acp{WER} of an automatic speech recognizer to evaluate the overall quality of the enhanced signals at the beamformer outputs, since a high speech recognition accuracy is the main goal in robot audition. As \ac{ASR} engine, we employed PocketSphinx \cite{Huggins:2006} with a \ac{HMM}-\ac{GMM}-based acoustic model which was trained on clean speech from the GRID corpus \cite{Cooke:2006}, using MFCC+$\Delta$+$\Delta \Delta$ features and cepstral mean normalization. For the computation of the \ac{WER} scores, only the letter and the number in the utterance were evaluated, as in the CHiME challenge \cite{ChristensenBMG10}. Our test signal contained $200$ utterances. Note that since the \ac{ASR} system was trained on clean speech, we implicitly measure the amount of target signal distortion and interferer suppression.

\begin{figure}
  \begin{tikzpicture}
    \begin{axis}[
      width=9.25cm,height=4cm,grid=major,grid style = {dotted,black},
      label style = {font=\scriptsize},
      tick label style = {font=\tiny},
      ylabel style={yshift=-2mm},
      ybar=0.5pt, 
      bar width=6pt,
      enlargelimits=0.075,
      ymin=0, ymax = 65,
      ytick={0,20,40,60},
      ylabel={$\text{WER}/\%\, \rightarrow$},
      xtick=data,
      xmin=0, xmax = 180,
      xlabel={$\phi_\text{ld}/\circ$},
      xlabel shift={-1mm},
      x dir=reverse,
      every node near coord/.append style={font=\tiny,
	  rotate=90, anchor=west,
	  /tikz/.cd},
      nodes near coords, nodes near coords align={vertical},
      legend style={at={(0.5,1.225)},anchor=north,legend columns=-1,font=\scriptsize},	
      legend entries={$\text{Input}\quad$, $\text{RLSFI}_\text{HRTF}\quad$, $\text{RLSFIP}_\text{HRTF}\quad$, $\text{RLSFIP}_\text{Free-field}$},
      ]	
      \addplot[black,fill=red,postaction={pattern=north east lines}] coordinates {(0,50.5) (30,48.9) (60,48.0) (90,47.1) (120,48.0) (150,48.6) (180,50.7)}; 
      \addplot[black,fill=green] coordinates {(0,33.7) (30,33.8) (60,32.1) (90,32.4) (120,34.5) (150,34.7) (180,35.7)};          
      \addplot[black,fill=yellow, postaction={pattern=crosshatch dots}] coordinates {(0,33.8) (30,36.3) (60,33.3) (90,32.5) (120,36.1) (150,37) (180,35.6)};
      \addplot[black,fill=cyan, postaction={pattern=north west lines}] coordinates {(0,46.9) (30,39.6) (60,38.2) (90,40.0) (120,39.7) (150,43.5) (180,44.1)};      
    \end{axis}
  \end{tikzpicture}
  \vspace{-4.5mm}
  \caption{Illustration of average target source position-specific \acp{WER} in $\%$, obtained at the input (red bars) and at the output of the \ac{HRTF}-based \ac{RLSFI} (green bars), \ac{HRTF}-based \ac{RLSFIP} (yellow bars), and free-field-based \ac{RLSFIP} (cyan bars) beamformers.}  
  \label{fig:results_WER}
  \vspace{-5mm}
\end{figure}
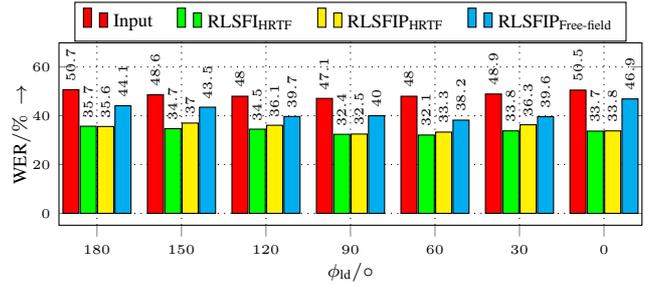

We evaluated the signal enhancement in a two-speaker scenario, where the target signal was located at positions between $\phi_\text{ld}=0^\circ$ and $\phi_\text{ld}=180^\circ$ in steps of $30^\circ$. The \ac{DOA} of the target signal was assumed to be known for the experiments, i.e., no localization algorithm was applied. An investigation of the \ac{HRTF}-based beamformer's robustness against \ac{DOA} estimation errors can be found in \cite{lnt2016-16}. For each target position, seven interfering speaker positions between $\phi_\text{int}=15^\circ$ and $\phi_\text{int}=165^\circ$ in steps of $30^\circ$ were evaluated. An overview over all source positions is given in Fig.~\ref{fig:evaluation_scenarios}, where target and interfering sources are represented by black circles and red crosses, respectively. 
We created the microphone signals by convolving clean speech signals with \acp{RIR} which we measured in a lab room with a reverberation time of $T_{60} \approx 190$ms and a critical distance \cite{kuttruff2000room} of approximately $1.2$m. The \acp{RIR} were measured with the same configuration as was used for the \ac{HRTF} measurements described above.
The \acp{WER} were calculated for each combination of target and interfering source position and averaged over the \acp{WER} obtained for the different interferer positions. The resulting average target source position-specific \acp{WER} are depicted in Fig.~\ref{fig:results_WER}.
The obtained \acp{WER} show that both \ac{HRTF}-based beamformers significantly improve the speech recognition accuracy of the input signal. Moreover, they also outperform the free-field-based \ac{RLSFIP} beamformer significantly, which emphasizes the necessity to incorporate the effect of the robot's head on the sound field into the beamformer design. It is interesting to see that the \ac{HRTF}-based \ac{RLSFIP} beamformer performs as well as the \ac{HRTF}-based \ac{RLSFI} beamformer whenever the target source is located in one of the \acp{PLD} which were used for designing the polynomial beamformer. When this is not the case, only a slightly higher average \ac{WER} is obtained. This confirms that the polynomial interpolation of the \ac{HRTF}-based \ac{RLSFI} beamformer design works reasonably well such that it can be used in a robot audition scenario.

\vspace{-3mm}
\section{Conclusion}
\label{sec:Conclusion}
In this work, we proposed an \ac{HRTF}-based \ac{RLSFIP} beamformer design which allows for a flexible steering of a previously proposed robust \ac{HRTF}-based \ac{RLSFI} beamformer.
We evaluated both beamformer designs with respect to their corresponding approximation error of the desired beamformer response and with respect to their signal enhancement performance which was evaluated by means of \acp{WER} of an \ac{ASR} system. The results showed that the polynomial beamformer design provides a good approximation of the \ac{RLSFI} beamformer design and, therefore, can be used successfully in a robot audition scenario instead of the computationally much more complex \ac{RLSFI} beamformer design.
Future work includes an investigation of the proposed \ac{HRTF}-based polynomial beamformer design for more irregular sensor arrangements
as well as an evaluation with a state-of-the-art \ac{DNN}-based \ac{ASR} system. 
An extension of the \ac{RLSFIP} beamformer design to allow for a flexible steering of the main beam in two dimensions is another aspect of future work.

\bibliographystyle{IEEEbib}
\bibliography{iwaenc2016}

\end{document}